\documentclass[twocolumn,showpacs,preprintnumbers,amsmath,aps]{revtex4}
\usepackage{graphicx}
\usepackage{dcolumn}
\usepackage{bm}
\begin{document}
\def\eq#1{(\ref{#1})}
\def\fig#1{{\ref{#1}}}
\title{Properties of the superconducting state in compressed Sulphur}
\author{A.P. Durajski, R. Szcz{\c{e}}{\`s}niak, M.W. Jarosik}
\affiliation{Institute of Physics, Cz{\c{e}}stochowa University of Technology, Al. Armii Krajowej 19, 42-200 Cz{\c{e}}stochowa, Poland}
\email{adurajski@wip.pcz.pl}
\date{\today} 
\begin{abstract}
The thermodynamic properties of the superconducting state in Sulphur under the pressure at $160$ GPa were determined. It has been shown that: 
(i) the critical value of the Coulomb pseudopotential is equal to $0.127$;
(ii) the critical temperature ($T_{C}=17$ K) should be calculated by using the modified Allen-Dynes formula; 
(iii) the effective electron-electron interaction is attractive in the range of frequencies from zero to the frequency slightly lesser than the maximum phonon frequency ($\sim 0.85\Omega_{{\rm max}}$);
(iv) the dimensionless ratios $2\Delta\left(0\right)/k_{B}T_{C}$, $\Delta C\left(T_{C}\right)/C^{N}\left(T_{C}\right)$ and  $T_{C}C^{N}\left(T_{C}\right)/H^{2}_{C}\left(0\right)$ are equal to $3.7$, $1.65$ and $0.16$ respectively; 
(v) the ratio of the effective to bare electron mass reaches maximum of $1.77$ for $T=T_{C}$. 
\end{abstract}
\maketitle
%
\section{Introduction}

At ambient pressure ($p$) Sulphur ({\it S}) is the insulator. Under compression {\it S} undergoes the sequence of structural phase transitions and metallizes \cite{Degtyareva}. At $93$ GPa, Sulphur is the superconductor with $T_{C}=10.1$ K \cite{Struzhkin}. The critical temperature increases linearly with the pressure up to $157$ GPa; the rate of $dT_{C}/dp$ is equal to $0.055$ ${\rm K/GPa}$ and $\left[T_{C}\right]_{p=157 {\rm GPa}}=14$ K. Near $160$ GPa, when {\it S} transforms to the $\beta$-Po-type phase, the critical temperature increases rapidly from $14$ K to $17$ K. We note that $T_{C}=17$ K is among the highest critical temperatures observed in the elemental solids (the maximum value of $T_{C}$, which is equal to $25$ K, has Calcium under the pressure at $161$ GPa \cite{Yabuuchi}).

In the paper we have studied the thermodynamic properties of the superconducting state in {\it S} for $p=160$ GPa. The numerical calculations have been conducted in the framework of the Eliashberg formalism \cite{Carbotte}.   

\section{The Eliashberg equations}
%
\begin{figure}[t]%
\includegraphics*[scale=0.31]{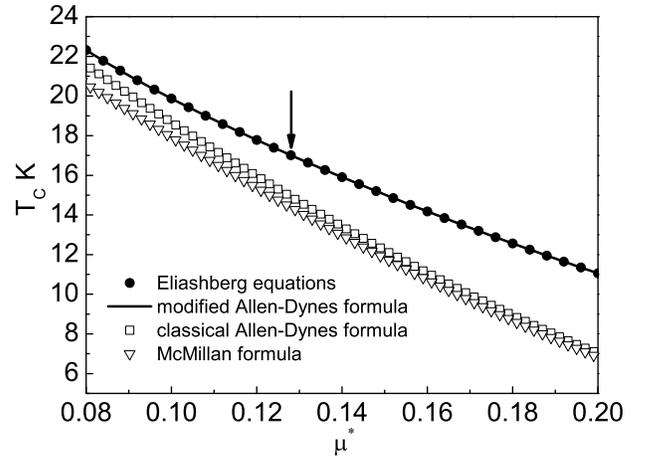}
\caption{The critical temperature as a function of $\mu^{*}$. The results have been obtained with an use of the following approaches: the Eliashberg equations, the modified (classical) Allen-Dynes formula and the McMillan expression. The arrow shows the experimental value of $T_{C}$ for $\mu_{C}^{*}=0.127$.}
\label{f1} 
\end{figure}
%

The Eliashberg set in the mixed representation takes the form:
\begin{eqnarray}
\label{r1}
\phi\left(\omega\right)&=&
\frac{\pi}{\beta}\sum_{m=-M}^{M}\frac{\left[\lambda\left(\omega-i\omega_{m}\right)-\mu^{*}\left(\omega_{m}\right)\right]}
{\sqrt{\omega_m^2Z^{2}_{m}+\phi^{2}_{m}}}\phi_{m}\\ \nonumber
                              &+& i\pi\int_{0}^{+\infty}d\omega^{'}\alpha^{2}F\left(\omega^{'}\right)
                                  [\left[N\left(\omega^{'}\right)+f\left(\omega^{'}-\omega\right)\right]\\ \nonumber
                              &\times&K\left(\omega,-\omega^{'}\right)\phi\left(\omega-\omega^{'}\right)]\\ \nonumber
                              &+& i\pi\int_{0}^{+\infty}d\omega^{'}\alpha^{2}F\left(\omega^{'}\right)
                                  [\left[N\left(\omega^{'}\right)+f\left(\omega^{'}+\omega\right)\right]\\ \nonumber
                              &\times&K\left(\omega,\omega^{'}\right)\phi\left(\omega+\omega^{'}\right)]
\end{eqnarray}
and
\begin{eqnarray}
\label{r2}
Z\left(\omega\right)&=&
                                  1+\frac{i\pi}{\omega\beta}\sum_{m=-M}^{M}
                                  \frac{\lambda\left(\omega-i\omega_{m}\right)\omega_{m}}{\sqrt{\omega_m^2Z^{2}_{m}+\phi^{2}_{m}}}Z_{m}\\ \nonumber
                              &+&\frac{i\pi}{\omega}\int_{0}^{+\infty}d\omega^{'}\alpha^{2}F\left(\omega^{'}\right)
                                  [\left[N\left(\omega^{'}\right)+f\left(\omega^{'}-\omega\right)\right]\\ \nonumber
                              &\times&K\left(\omega,-\omega^{'}\right)\left(\omega-\omega^{'}\right)Z\left(\omega-\omega^{'}\right)]\\ \nonumber
                              &+&\frac{i\pi}{\omega}\int_{0}^{+\infty}d\omega^{'}\alpha^{2}F\left(\omega^{'}\right)
                                  [\left[N\left(\omega^{'}\right)+f\left(\omega^{'}+\omega\right)\right]\\ \nonumber
                              &\times&K\left(\omega,\omega^{'}\right)\left(\omega+\omega^{'}\right)Z\left(\omega+\omega^{'}\right)], 
\end{eqnarray}
where:
\begin{equation}
\label{r3}
K\left(\omega,\omega^{'}\right)\equiv
\frac{1}{\sqrt{\left(\omega+\omega^{'}\right)^{2}Z^{2}\left(\omega+\omega^{'}\right)-\phi^{2}\left(\omega+\omega^{'}\right)}}.
\end{equation}

The symbols $\phi\left(\omega\right)$ and $\phi_{m}\equiv\phi\left(i\omega_{m}\right)$ denote the order parameter functions on the real and imaginary axis respectively; $Z\left(\omega\right)$ and $Z_{m}\equiv Z\left(i\omega_{m}\right)$ are the wave function renormalization factors; $m$-th Matsubara frequency is given by: $\omega_{m}\equiv \left(\pi / \beta\right)\left(2m-1\right)$, where $\beta\equiv\left(k_{B}T\right)^{-1}$ and $k_{B}$ is the Boltzmann constant. In the framework of the Eliashberg formalism, the order parameter is defined as: $\Delta\equiv \phi/Z$. The electron-phonon pairing kernel has the form: $\lambda\left(z\right)\equiv 2\int_0^{\Omega_{\rm{max}}}d\Omega\frac{\Omega}{\Omega ^2-z^{2}}\alpha^{2}F\left(\Omega\right)$, where
the Eliashberg function for Sulphur under the pressure at $160$ GPa ($\alpha^{2}F\left(\Omega\right)$) has been calculated in the paper \cite{Rudin}; the maximum phonon frequency ($\Omega_{\rm{max}}$) is equal to $86.7$ meV.
The function $\mu^{*}\left(\omega_{m}\right)\equiv\mu^{*}\theta\left(\omega_{c}-|\omega_{m}|\right)$ describes the electron depairing interaction;  $\theta$ denotes the Heaviside unit function and $\omega_{c}$ is the cut-off frequency ($\omega_{c}=3\Omega_{\rm{max}}$). 
The critical value of the Coulomb pseudopotential ($\mu^{*}_{C}$) should be calculated by using the condition: $\Delta_{m=1}\left(\mu^{*}\right)=0$ for $T=T_{C}$. The following result has been obtained: $\mu^{*}_{C}=0.127$.  
The symbols $N\left(\omega\right)$ and $f\left(\omega\right)$ denote the statistical functions of bosons and fermions respectively.

The Eliashberg equations have been solved for $2201$ Matsubara frequencies ($M=1100$) by using the numerical method presented in the paper \cite{Szczesniak1}. In the considered case the functions $\phi$ and $Z$ are stable for $T\geq 4$ K.

\section{Results}

%
\begin{figure}[t]%
\includegraphics*[scale=0.31]{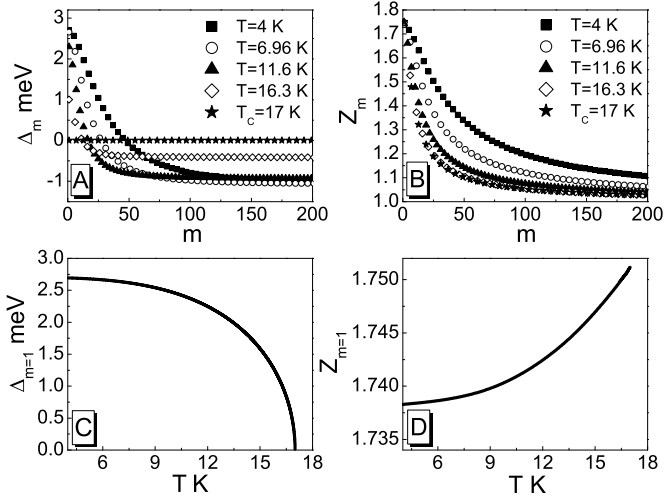}
\caption{(A)-(B) The order parameter and the wave function renormalization factor on the imaginary axis for the selected temperatures. (C)-(D) The dependence of $\Delta_{m=1}$ and $Z_{m=1}$ on the temperature. The values of $\Delta_{m=1}\left(T\right)$ can be fitted by the expression:  $\Delta_{m=1}\left(T\right)=\Delta_{m=1}\left(0\right)\sqrt{1-\left(\frac{T}{T_{C}}\right)^{\beta}}$, where $\Delta_{m=1}\left(0\right)=2.7$ meV and $\beta=3.37$.}
\label{f2}
\end{figure}
%
%
\begin{figure}[t]%
\includegraphics*[scale=0.31]{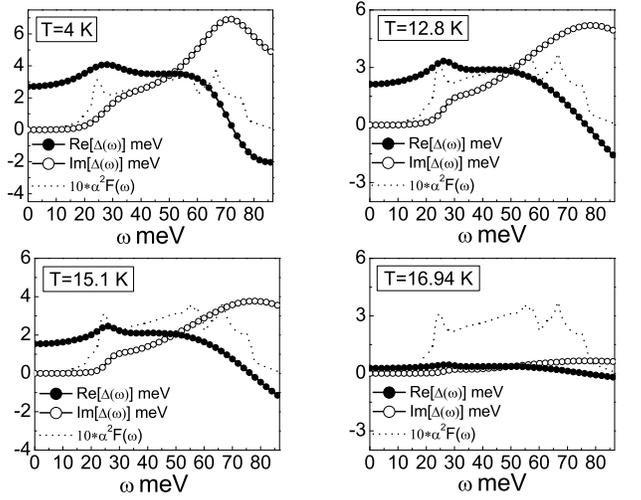}
\caption{
The dependence of the real and imaginary part of the order parameter on the frequency for selected temperatures. The rescaled Eliashberg function is also plotted.} 
\label{f3}
\end{figure}
%
\begin{figure}[t]%
\includegraphics*[scale=0.33]{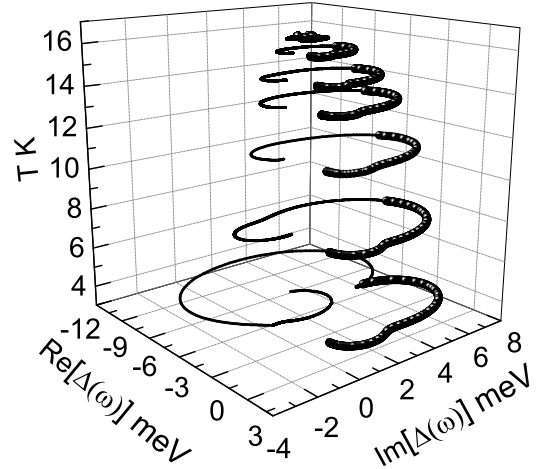}
\caption{The order parameter on the complex plane for the selected values of the temperature. The bold lines represent the solutions for $\omega\in\left<0,\Omega_{\rm max}\right>$; whereas regular lines correspond to the solutions for $\omega\in\left(\Omega_{\rm max}, \omega_{c}\right>$.} 
\label{f4}
\end{figure}
%


The classical Allen-Dynes or McMillan formula cannot be used in the case of Sulphur (see Fig. \fig{f1}) \cite{AllenDynes}, \cite{McMillan}. For this reason, we have obtained modified Allen-Dynes expression on the basis of $330$ exact values of $T_{C}\left(\mu^{*}\right)$. In particular:
\begin{equation}
\label{r4}
k_{B}T_{C}=f_{1}f_{2}\frac{\omega_{\rm ln}}{1.45}\exp\left[\frac{-1.03\left(1+\lambda\right)}{\lambda-\mu^{*}\left(1+0.06\lambda\right)}\right],
\end{equation}
where $f_{1}$ is the strong-coupling correction function ($f_{1}\equiv\left[1+\left(\frac{\lambda}{\Lambda_{1}}\right)^{\frac{3}{2}}\right]^{\frac{1}{3}}$) and $f_{2}$ denotes the shape correction function 
($f_{2}\equiv 1+\frac{\left(\frac{\sqrt{\omega_{2}}}{\omega_{\rm{ln}}}-1\right)\lambda^{2}}{\lambda^{2}+\Lambda^{2}_{2}}$). The symbol $\omega_{2}$ denotes the second moment of the normalized weight function, $\omega_{{\rm ln}}$ is the logarithmic phonon frequency and $\lambda$ is called the electron-phonon coupling constant. For Sulphur the parameters $\sqrt{\omega_{2}}$, $\omega_{{\rm ln}}$ and $\lambda$ are equal to $45.39$ meV, $37.7$ meV and $0.75$ respectively. The functions $\Lambda_{1}$ and $\Lambda_{2}$ have the form:
$\Lambda_{1}\equiv 55\mu^{*}-1$ and $\Lambda_{2}\equiv 11.06\mu^{*}\left(\frac{\sqrt{\omega_{2}}}{\omega_{\rm{ln}}}\right)$.


In Figs. \fig{f2} (A) and (B) we have shown the solutions of the Eliashberg equations on the imaginary axis for the temperature range from $4$ K to $17$ K. Additionally, in Figs. \fig{f2} (C) and (D) the functions $\Delta_{m=1}\left(T\right)$ and $Z_{m=1}\left(T\right)$ are plotted. On the basis of the presented results one can state that the value of the order parameter at the temperature of zero Kelvin  ($\Delta\left(0\right)$) is equal to $\Delta\left(T=4{\rm K}\right)$ with the good approximation. In order to calculate $\Delta\left(T=4{\rm K}\right)$ the following algebraic equation has to be used: 
$\Delta\left(T\right)={\rm Re}\left[\Delta\left(\omega=\Delta\left(T\right)\right)\right]$. The form of the order parameter on the real axis for $T=4$ K is presented in Fig. \fig{f3} (A). With the help of the simple calculation one can obtain: $\Delta\left(T=4{\rm K}\right)=2.71$ meV. Then, the value of the ratio $R_{1}\equiv\frac{2\Delta\left(0\right)}{k_{B}T_{C}}$ is equal to $3.7$. We notice, that in the framework of the BCS model, the parameter $R_{1}$ takes the lower value: $\left[R_{1}\right]_{{\rm BCS}}=3.53$ \cite{BCS}.

On the basis of Figs. \fig{f3} (A)-(D) we have stated that Re$\left[\Delta\left(\omega\right)\right]$ and Im$\left[\Delta\left(\omega\right)\right]$ are plainly correlated with the shape of the Eliashberg function in the full range of the considered temperatures. From the physical point of view this indicates, that the form of order parameter on the real axis clearly reflects the form of the electron-phonon interaction in Sulphur. Additionally, in Fig. \fig{f4} we have plotted the order parameter on the complex plane for the selected temperatures; the frequencies from $0$ to $\omega_{c}$ have been taken into consideration. We have found that the values of $\Delta\left(\omega\right)$ form the distorted spiral with the radius
that decreases together with the temperature growth. One can also see, that the effective electron-electron interaction is attractive (Re$\left[\Delta\left(\omega\right)\right] > 0$) in the range of the frequencies from zero to $\sim 0.85\Omega_{{\rm max}}$.     

\begin{figure}[t]%
\includegraphics*[scale=0.31]{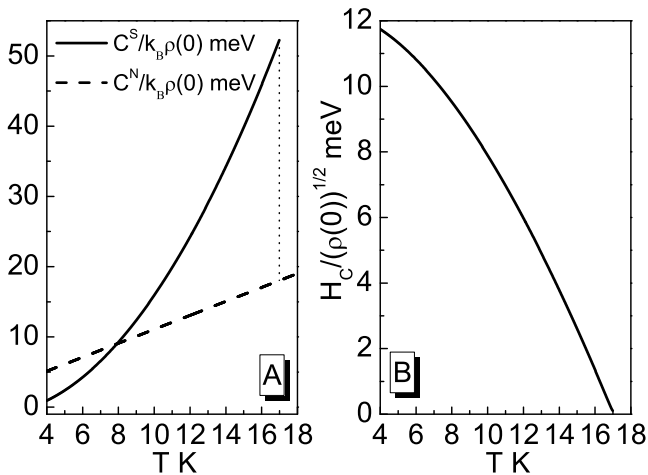}
\caption{(A) The dependence of the specific heat in the superconducting and normal state on the temperature. (B) The thermodynamic critical field as a function of the temperature.}
\label{f5}
\end{figure}
%


The free energy difference between the superconducting and normal state ($\Delta F$) can be calculated on the basis of the formula \cite{Carbotte}:
$\frac{\Delta F}{\rho\left(0\right)}=-\frac{2\pi}{\beta}\sum_{m=1}^{M}\left(\sqrt{\omega^{2}_{m}+\Delta^{2}_{m}}- \left|\omega_{m}\right|\right)
(Z^{{\rm S}}_{m}-Z^{N}_{m}\frac{\left|\omega_{m}\right|}{\sqrt{\omega^{2}_{m}+\Delta^{2}_{m}}})$, where $\rho\left(0\right)$ denotes the value of the electronic density of states at the Fermi energy. The symbols $Z^{S}_{m}$ and $Z^{N}_{m}$ represent the wave function renormalization factors for the superconducting (S) and normal (N) state respectively. 

With the help of $\Delta F$ the specific heat in the superconducting ($C^{S}$) and normal ($C^{N}$) state, as well as, the thermodynamic critical field ($H_{C}$) have been determined \cite{Carbotte}. In Fig. \ref{f5} (A)-(B) we have presented the dependence of the specific heats and the thermodynamic critical field on the temperature. On the basis of determined thermodynamic functions the values of the ratios $R_{2}\equiv \Delta C\left(T_{C}\right)/C^{N}\left(T_{C}\right)$ and $R_{3}\equiv T_{C}C^{N}\left(T_{C}\right)/H^{2}_{C}\left(0\right)$ have been calculated. We have obtained: $R_{2}=1.65$ and $R_{3}=0.16$. We notice, that the BCS model predicts: $\left[R_{2}\right]_{\rm BCS}=1.43$ and $\left[R_{3}\right]_{\rm BCS}=0.168$ \cite{BCS}. It is easy to see, that for Sulphur, $R_{2}$ and $R_{3}$ differ from the BCS values. 

Finally, we have calculated the ratio of the electron effective mass ($m^{*}_{e}$) to the bare electron mass ($m_{e}$):
$m^{*}_{e}/m_{e}={\rm Re}\left[Z\left(0\right)\right]$. It has been stated, that $m^{*}_{e}$ is relatively high in the full range of the considered temperatures and $\left[m^{*}_{e}/m_{e}\right]_{\rm max}=1.77$ for $T=T_{C}$.

\section{Summary}

The high-pressure superconducting state in Sulphur have been analyzed in the framework of the Eliashberg approach. It has been stated that the exact values of the thermodynamic parameters cannot be calculated by using of the simple BCS model. 

%

%
\end{document}